# Excitation of two-photon photoemission where epsilon is near zero on Ag(111)

Marcel Reutzel,[1,‡] Andi Li,[1] Branko Gumhalter,[2] and Hrvoje Petek[1,‡]

[1]Department of Physics and Astronomy and Pittsburgh Quantum Institute, University of Pittsburgh, Pittsburgh, Pennsylvania 15260, USA

[2]Institute of Physics, HR-10000 Zagreb, Croatia

email: mar331@pitt.edu, petek@pitt.edu

Abstract.

Because silver has epsilon-near-zero (ENZ) in the near-UV spectral region, its optical response is expected to be dominantly plasmonic and nonperturbatively nonlinear. The ENZ properties of Ag have made it the material of choice for investigation of plasmonic optical effects in physics, chemistry, optics, and nanotechnology. We investigate the nonlinear angle-resolved two-photon photoemission (2PP) spectroscopy of Ag(111) surface in the ENZ region. In addition to the well-understood and documented spectroscopic features of Einsteinian photoemission involving dipole excitations of the single-particle occupied and unoccupied surface states, 2PP spectra possess other distinct features where the optical fields excite or are modified by the collective virtual and real plasmon excitations.

The coherent electromagnetic response of a metal is described by its complex dielectric tensor $\varepsilon(\omega)$. The electronic time scale for optical excitation is dominated by the bulk plasma response, which is resonant at the frequency where $\varepsilon(\omega) \approx 0$. Below the plasma frequency, the electric potentials are nonlocal and retarded because the virtual collective plasmon response screens the optical field as well as the consequent Coulomb fields created by optical dipole excitations. Above $\varepsilon(\omega) \approx 0$, the electromagnetic fields propagate through metals as complex charge density fluctuations (plasmons), which can decay into single particle excitations [1-7]. Although angle-resolved photoemission spectra (ARPES) measure energy and momentum distributions of single photoelectrons [8], they carry information on both the electronic band-structures as well as the many-body processes that are turned on by photoexcitation [9-11]. When the photon frequency exceeds that of the bulk plasmon, the many-body response can produce spectroscopic features, which are referred to as "the final state effects", like plasmon satellite peaks that decorate main photoemission features in x-ray photoelectron spectroscopy [8]. The time scales for the prompt photoemission processes and their collective echoes can nowadays be measured by attosecond streaking experiments [12]. Much less is known, however, about how screening of optical fields by the retarded, nonlocal plasmon response affects photoemission spectra below the bulk plasma frequency [13-17]. Influence of the many-body response to optical fields on photoemission can be elucidated by scanning the photon energy through the epsilon-near-zero (ENZ) region ($\varepsilon(\omega) \approx 0$) [18,19].

The frequency dependence of plasmonic contributions to photoemission spectra has been observed and interpreted for free electron metals such as Al and Be [14,16,19], whose plasma frequencies exceed their work functions; this is difficult to achieve for one of the most intensively investigated plasmonic materials, Ag, because interband excitations reduce its plasmon frequency to $\hbar\omega_p \approx$ 3.8-3.9 eV, which is below the Ag work function of $\approx$4.5 eV [20,21]. Barman *et al.* demonstrated that virtual and real plasmonic responses strongly influence photoemission from Ag surfaces by using chemisorption of alkali atoms to reduce their work functions to below $\hbar\omega_p$ [18].

In this Letter, we study the nonlinear electronic responses of the pristine Ag(111) surface by two-photon photoemission (2PP) spectroscopy [22], a technique which is particularly sensitive to surface fields because it is nonlinear and photoelectron escape depths are a few nm, when tuning $\hbar\omega$ through the ENZ region. In Fig. 1(a), we plot the experimentally recorded real and the imaginary parts of $\varepsilon(\omega)$ for polycrystalline Ag (adopted from Ref. [23]), which describe the frequency dependent bulk response to an external electric field $E(\omega)$. The transient coherent polarization induced in the sample is $P(\omega) = \sum_{l=1}^{\infty} l\epsilon_0\chi_l E^l(\omega)$, where $\chi(\omega) = \varepsilon(\omega) - 1$ ($\varepsilon_0$: vacuum permittivity, $\chi_l$: susceptibility of the order, *l*). For Ag, the ENZ condition occurs when $\hbar\omega \approx$ 3.8-3.9 eV, through cancellation of the negative and positive contributions to Re[$\varepsilon(\omega)$], respectively, from the Drude response of free electrons and bulk interband transitions [24], while Im[$\varepsilon(\omega)$] is still relatively small [mainly contributed by intraband absorption; inset Fig. 1(a)]. The field experienced by electrons near a metal surface is given by how its q = 0 response function modifies the external field [11]. The response function has real and imaginary parts, which depend on $\omega$ and are contributed by various transitions, which total q = 0. The imaginary part is an on-the-energy-shell quantity describing real transitions; for example, its bulk plasmon component is zero for $\hbar\omega < \hbar\omega_p$, and analogously for other, *e.g.*, interband excitations. In other words, it rises stepwise as $\hbar\omega$ crosses $\hbar\omega_p$. Thus, at ENZ, the plasmonic properties dominate the optical responses of metals [25,26] and optical reflection is minimum due to the excitation of bulk plasmons [24]. Furthermore, the optical response is *nonperturbative*, because the linear term describing it is $P^{(l=1)}(\omega) \approx 0$, and therefore higher-order nonlinear terms dominate [27-29], leading to, for example, strong enhancement of surface second-harmonic generation [30]. Below ENZ, external fields are strongly screened by virtual plasmon excitation and above ENZ they are transmitted through the crystal by charge density fluctuations (or plasmons). In addition, the surface field has an intensity maximum below ENZ, because of the nonlocal multipole plasmon response, and above ENZ, it has another maximum brought about by the bulk plasmon excitations [19,31].

We perform angle-resolved 2PP spectroscopy of Ag(111) surface at 90 K using femtosecond, $p$-polarized, laser pulses in the $2.6 < \hbar\omega < 4.5$ eV range that encompasses the ENZ region. The excitation light is generated by a Clark MXR Impulse fiber laser oscillator-amplifier, which excites a noncollinear optical parametric amplifier (NOPA) at a 1 MHz repetition rate. One color 2PP is excited by frequency doubled output of the NOPA with an average power of ≈1-10 mW, and pulse duration of ≈20-30 fs. The optical and vacuum systems, as well as the sample preparation, are described in Refs. [32-34]. Energy- and $k_\|$-momentum-resolved 2PP spectra for different $\hbar\omega$ and their line profiles for $k_\| = 0$ are shown in Figs. 2 and 3, respectively; the photoelectron energies, $E_f$, relative to the Fermi energy, $E_F$, vs. $\hbar\omega$ are plotted in Fig. 4(a) for various 2PP spectral features.

Figure 2(a) shows an $E_f(k_\|)$ resolved 2PP spectrum excited with $\hbar\omega = 3.32$ eV, which is similar to previously reported 2PP spectra that have been excited with frequency doubled Ti:Sapphire lasers [33,35-37]; the spectrum includes signal from a nonresonant two-photon absorption from the Shockley surface state (SS) and a two-photon resonant transition from the lower, $L_{sp}$, to the upper, $U_{sp}$, bulk $sp$-bands [SP transition; see the excitation diagram in Fig. 3(a)]. The $k_\|$ for the SP transition spans the accessible range, whereas for SS it is limited by occupation to where SS disperses above $E_F$ [32]. Increasing $\hbar\omega$ shifts the SS and SP transitions to higher $E_f$ [Fig. 2(b), Fig. 3(b)]. The resonant $n = 1$ image potential state (IP) ← SS transition occurs at $\hbar\omega \approx 3.9$ eV, but resonance enhancement occurs only for the occupied $k_\parallel$-range of SS [Fig. 2(c)]. The $E_f$ of the SS and IP state peaks tune with $2\hbar\omega$ and $1\hbar\omega$, respectively, so for nonresonant excitation ($\hbar\omega \lesssim 3.9$ eV) they are detected separately [Fig. 4(a)]. 2PP spectroscopy of the SS and IP states has been investigated extensively [32,33,35-42], typically using a two-color UV and IR excitation schemes. Neither state appears to be obviously sensitive to the ENZ condition, although that could be masked by a fortuitous near coincidence of their resonance with $\hbar\omega_p \approx 3.8$-3.9 eV. Instead, we describe how the SP transition and a spectral feature at $E_f \approx 7.75$ eV (Fig. 4) reveal the multipole plasmon ($\hbar\omega < \hbar\omega_p$) and bulk plasmon ($\hbar\omega \geq \hbar\omega_p$) responses, respectively.

First, we consider the resonant two-photon SP transition between the bulk $sp$-bands, which form the dominant spectroscopic feature below ENZ for excitation with $\hbar\omega < \hbar\omega_p$. Although, the SP transition can be excited in the entire $\hbar\omega$ range, its intensity in 2PP varies drastically: it is strongest for $\hbar\omega \approx 3.4$-3.5 eV, where it dominates the spectra, but it disappears above $\hbar\omega_p$ [Fig. 4(b)]. Such intensity variation cannot be attributed to transition moments, because it does not exist in one-photon photoemission (1PP), when $\hbar\omega = 6$-10 eV [43] is used to excite the same initial and final states. Instead, 2PP being proportional to $E(\omega)^4$ is highly sensitive to electric field strength at the surface. Strong modulation of the total surface field is expected below $\hbar\omega_p$ because the optical field is modulated by the surface dielectric, *i.e.*, multipole plasmon, response [16,17,19]. The multipole plasmon frequency of Ag(111) has been difficult to define because it nearly overlaps with the surface plasmon resonance, but Rocca and coworkers reported it to be $\hbar\omega_{mp} = 3.74$ eV based on electron energy loss spectroscopy (EELS) [44]. Our finding of the SP transition enhancement maximum below their $\hbar\omega_{mp}$ is consistent with the multipole plasmon enhancement of surface fields, which occurs over a broad frequency range [17,45,46], but it should not be considered to directly measure $\hbar\omega_{mp}$, because other factors can also influence the 2PP signal. We note, that the band structures, electron escape depths, *etc.* that affect 2PP spectra of Cu(111) and Ag(111) surfaces are nearly identical, except for the plasmonic response [7]. This may explain why the corresponding SP transition in Cu(111) is barely detected in 2PP spectra (see supplemental material S1, Fig. S1 and Ref. [47]), whereas for Ag(111) it dominates, but in a narrow range ($\hbar\omega < \hbar\omega_p$).

Next, we consider the spectral feature at $E_f \approx 7.75$ eV $\approx 2\hbar\omega_p$, which appears for excitation with $\hbar\omega \geq 3.9$ eV; as shown in Fig. 2(c)-(d), its photoemission line shape is asymmetric and is detected over the accessible $k_\parallel$-range. As is evident in Fig. 3(c) and Fig. 4(a), the $2\hbar\omega_p$-feature does not tune with $\hbar\omega$, which could imply either that it is localized by excitation of a discrete final state, or else that it involves a collective excitation. The $2\hbar\omega_p$-feature was only reported by Giesen *et al.* using nanosecond laser, one-color UV pulse excitation [38]; they excluded assignment to a final state effect because only $U_{sp}$ exits and its parabolic dispersion does not localize photoemission at $E_f \approx 7.75$ eV (cf. band-diagram in Fig. 3(a) and Ref. [18,43,48]). Instead, they attributed it to an Auger process, where two electrons excited to the $n = 1$ IP state interact, causing one to decay to $E_F$ and the other to be emitted at $E_f \approx 7.8$ eV to conserve energy. Because the $n = 1$ IP state lifetime [41] is comparable to our laser pulse duration, which is six orders-of-magnitude shorter than that of Giesen *et al.*, the putative Auger process of IP state electrons is unlikely to be observed with comparable relative intensity to the IP state in both experiments (see as well supplemental material S2, Fig. S3). In addition, we exclude the Auger process by deposition of organic molecules onto the Ag(111) surface [49], which quenches the surface SS and $n = 1$ IP state spectral contributions differently from the $2\hbar\omega_p$-feature, indicating that they are unrelated (supplemental material S3, Fig. S4). Instead, we attribute the $2\hbar\omega_p$-feature to a process beyond Einsteinian photoemission, specifically, the decay of two bulk plasmon quanta causing excitation of single electrons from $E_F$, as we outline in detail.

We assign the $2\hbar\omega_p$-signal to a process beyond the single-particle band structure of Ag(111) that turns on abruptly at $\hbar\omega \geq \hbar\omega_p$, i.e., the onset of excitations of bulk plasmons at $\varepsilon(\omega) = 0$, which is detected by their decay into single particle excitations. For this process to appear as a peak, however, the decay must selectively excite single electrons from $E_F$ to excite photoelectrons to $E_f = 2\times\hbar\omega_p + E_F$. That such excitation occurs is plausible because constant initial state 1PP spectra, monitoring photoemission yield from $E_F$, have been found to be strongly enhanced at $\hbar\omega_p$ [16,18]. Also, an EELS spectral peak of Ag(111) at 7.6 eV has been assigned to a two-quantum $\hbar\omega_p$ loss [50].

Our assignment of the $2\hbar\omega_p$-feature in 2PP spectra is supported by its abrupt threshold at $\hbar\omega \geq \hbar\omega_p$, the critical condition for excitation of real bulk plasmons. Moreover, its constant $E_f$ implies a non-Einsteinain photoemission process where the photoelectron energy is independent of $\hbar\omega$. We note that in 2PP we cannot observe a lower-order bulk plasmon decay, i.e. a $1\hbar\omega_p$-feature at $E_f \approx 3.8$ eV, because the generated hot electrons could not overcome the work function of Ag(111). Chemisorption of Rb atoms onto Ag(111) to lower its work function with negligible perturbation to Ag band structure [51-55] enables our access to such hot electrons. Indeed, 1PP spectra of Rb/Ag(111) with Hg-lamp excitation ($\hbar\omega \approx 4.86$ eV; supplemental material S4, Fig. S5) reveal, in addition to the expected single particle features, a broad peak at $E_f \approx 3.7$ eV that could be caused by $1\hbar\omega_p$ decay. Also, a shoulder at $E_f \approx 3.7$ eV appears in 1PP spectra of Na/Ag(100) by Barman *et al.* [18]. Thus, we attribute the $1\hbar\omega_p$-feature to decay bulk plasmons exciting electrons from $E_F$.

In contrast to Einsteinian photoemission by optical fields, the $2\hbar\omega_p$-feature cannot be interpreted as the bulk plasmon acting as an optical field to excite dipole transitions. The surface projected single-particle band structure of Ag(111) has a band gap at $k_\parallel = 0$ that extends from $E-E_F$ = -0.4 to 3.9 eV [56], and thus no electrons can be excited from $E_F$ by a single particle excitation. The SS state, which is very close to $E_F$, also cannot be the initial state for such a process, because (*i*) its $k_\parallel$-dispersion is narrower and different from the $2\hbar\omega_p$-feature [Fig. 2(e)], and (*ii*) it is quenched more rapidly by molecular adsorption (Fig. S4). Furthermore, we verify that the $2\hbar\omega_p$-feature involves excitation of electrons at $E_F$ by measuring the temperature dependence of 2PP spectra (supplemental material S5, Fig. S6). In 2PP spectra taken at 90 and 300 K, the $2\hbar\omega_p$-feature is strongly temperature dependent as expected for electrons near $E_F$, which are most sensitive to the Fermi-Dirac occupation function and electron-phonon interactions. Even though $E_F$ ($k_\parallel = 0$) lies within the band gap in the single-particle band structure, the decay of a collective excitation can produce the observed energy and momentum distribution of the $2\hbar\omega_p$-feature. Electron charge-density fluctuations, which constitute a bulk plasmon, occur at $E_F$, so it is conceivable that the same electron population is also involved in its decay; the electron photodynamics that create the $2\hbar\omega_p$-feature thus cannot be explained within the single-particle band structure of Ag(111).

We have investigated 2PP spectroscopy of pristine Ag(111) in the near-UV spectral region when tuning the photon energy through ENZ: For the SP transition and electrons emitted to $E_f = 2\hbar\omega_p$, photoemission is not directly excited by the optical field but involves intermediary plasmonic responses. The 2PP spectra of Ag(111) thus have contributions from the single-particle surface and bulk states as well as collective plasmonic excitations. The extensively studied surface state spectra do not appear to be strongly influenced by the plasmonic responses. By contrast, the two-photon resonant excitation between the *sp*-bands of Ag has a pronounced intensity variation that does not exist in 1PP spectra. Because nonlinear processes are enhanced at ENZ [28], and 2PP is surface sensitive method, such enhancement for Ag(111) can occur through the multipole plasmon resonance, which represents the surface dielectric response [15]. The scenario for the intermediation of 2PP by the Mie plasmons field has been described by Pfeiffer and coworkers [57] and a general description of how dielectric screening generates local fields that affect nonlinear electromagnetic processes has been developed by Timm and Bennemann [10]. Consistent with these predictions, we observe that screening of the optical field by the virtual plasmon excitation strongly modulates 2PP intensities below the bulk plasmon resonance. Furthermore, above the bulk plasmon resonance, we find that a multi-quantum decay of on-the-energy-shell plasmon excitations generates features in 2PP spectra. The signature of this process is the $\hbar\omega$-independent nonlinear electron emission at $E_f = 2\hbar\omega_p$. Similar nonlinear plasmon-induced photoemission has recently been invoked in space- and time- resolved photoemission electron microscopy [58] of plasmonic nanostructures [59]. Therefore, in the ENZ region, we find signatures of non-Einsteinian photoemission where photoelectron distributions are not defined only by the external optical field and the single-particle band structures, but also include contributions from the intermediate collective plasmonic excitations. These plasmonic responses strongly modulate the local fields, as evident in the consequent nonlinear photoemission intensities, and can even generate photoemission features beyond the single-particle band structures. Our findings are highly significant as they demonstrate how virtual and real collective excitations can enhance the local fields in the ENZ region and thereby affect nonlinear optical processes as well as generation of hot electrons in plasmonically enhanced optical processes.

**Acknowledgements**

The authors gratefully acknowledge support from DOE-BES Division of Chemical Sciences, Geosciences, and Biosciences Grant No. DE-SC0002313, and M. R. acknowledges support from the Alexander von Humboldt Foundation within the Feodor Lynen Fellowship Program. The authors thank Thomas Fauster and Karsten Horn for discerning discussions and sharing of unpublished data.

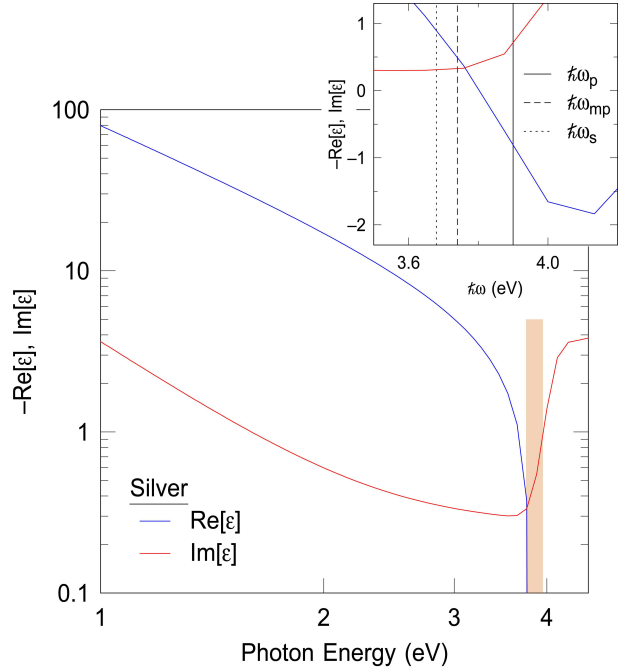

**Figure 1** | Frequency dependent dielectric function $\varepsilon(\omega)$ of silver plotted on a double-logarithmic scale as adapted from Ref. [23]; the brown shading highlights the energy region of ENZ. The inset shows a blow-up of Re[$\varepsilon(\omega)$] and Im[$\varepsilon(\omega)$] around ENZ on a linear scale; the bulk plasmon frequency $\hbar\omega_p$ as well as the monopole, $\hbar\omega_s$, and the multipole, $\hbar\omega_{mp}$, surface plasmon frequencies are indicated by vertical lines, the frequencies are adopted from Ref. [18,44]. In the ENZ region, the Im[$\varepsilon(\omega)$] is near a minimum where Re[$\varepsilon(\omega)$] passes through zero.

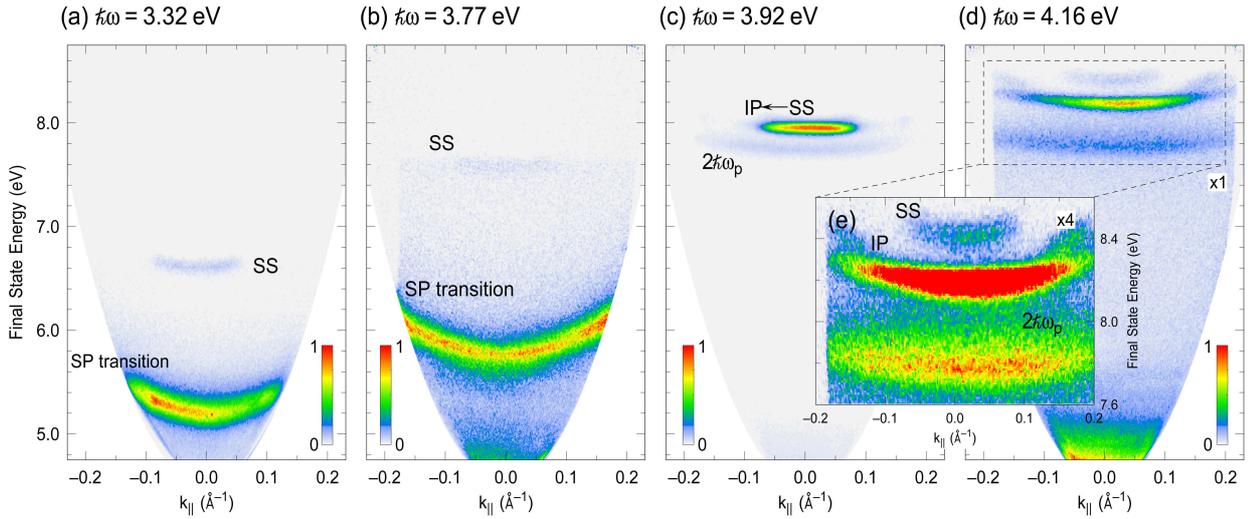

**Figure 2** | Energy and $k_\parallel$-momentum resolved 2PP spectra for different excitation energies $\hbar\omega$ increasing from (a) to (d), each color table is scaled separately; the main spectral features are labelled in each spectrum. (a)/(b) For $\hbar\omega < \hbar\omega_p$, the SP transition intensity dominates the SS state. (c)/(d) For $\hbar\omega \geq \hbar\omega_p$, the SS and the $n = 1$ IP states are detected, but not the SP transition. An additional feature, which cannot be assigned to the single particle band structure of Ag(111), is detected at $E_f = 2\hbar\omega_p \approx$ 7.75 eV. (e) Expanded and enhanced $E_f(k_\parallel)$-spectra from within the dashed box in (d) (the color-scale is multiplied by a factor of 4).

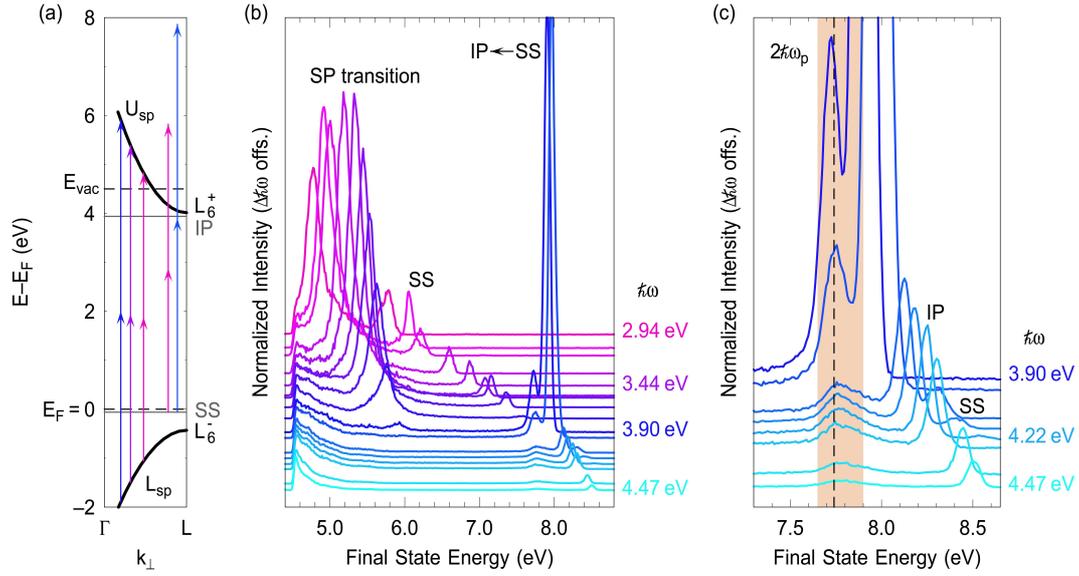

**Figure 3** | 2PP spectra of Ag(111) in a photon energy range from 2.9 to 4.5 eV ($k_\parallel = 0$) that have been normalized at the work function edge; the spectra are shifted vertically by the photon energy difference. (a) The projected single-particle band structure of Ag along the Γ-L direction [$k_\perp$ direction of Ag(111)]; the parabolic dispersion of the lower, $L_{sp}$, and the upper, $U_{sp}$, sp-bands as well as the surface states (SS, IP) that do not disperse with $k_\perp$, are labelled. Colored arrows indicate selected excitation pathways that contribute to 2PP spectra. (b) Photon energy dependent 2PP spectra; the main features are labelled in the figure. (c) Expanded 2PP spectra, the asymmetric peak at $2\hbar\omega_p \approx 7.75$ eV, which is highlighted by the brown box, has a constant $E_f$ for increasing $\hbar\omega$, and cannot be attributed to the band structure in (a).

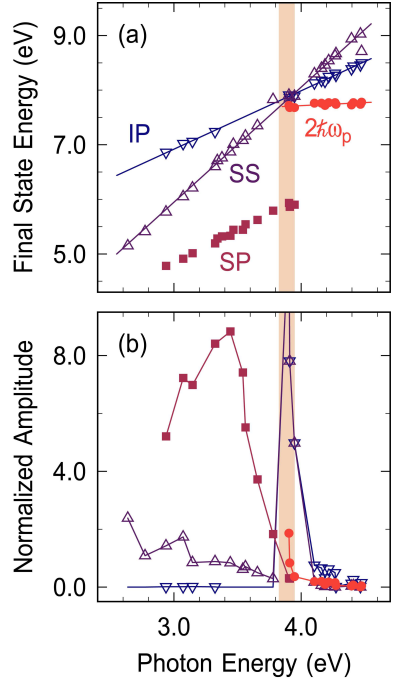

**Figure 4** | Quantitative evaluation of 2PP spectra of Ag(111) from Fig. 3 for $2.6 < \hbar\omega < 4.5$ eV ($k_\parallel = 0$); the brown shading highlights the energy region that corresponds to ENZ. (a) $E_f$ vs. $\hbar\omega$ for the major spectroscopic features. Slopes of 1 and 2, indicating one- or two-photon photoemission, are extracted for the intermediate IP and the initial SS state, respectively. The $2\hbar\omega_p$-feature is only observed for $\hbar\omega \geq 3.9$ eV and does not shift with $\hbar\omega$. The $\hbar\omega$ vs. $E_f$ dependence of the SP transition (initial and final sp-band states) is discussed in Ref. [35]. (b) Peak amplitudes vs. $\hbar\omega$: the SP transition amplitude peaks at $\hbar\omega \approx 3.4$-$3.5$ eV, significantly decreases towards $\hbar\omega \approx 3.9$ eV, and vanishes above it. Resonant excitation of $n = 1$ IP ← SS results in a strong enhancement of photoemission intensity at $\hbar\omega = 3.92$ eV. The markers indicate the same states/transitions as labelled in (b).

# References.

**Supplemental Material**

# Excitation of two-photon photoemission where epsilon is near zero on Ag(111)

Marcel Reutzel,[1,‡] Andi Li,[1] Branko Gumhalter,[2] and Hrvoje Petek[1,‡]

[1]*Department of Physics and Astronomy and Pittsburgh Quantum Institute, University of Pittsburgh, Pittsburgh, Pennsylvania 15260, USA*

[2]*Institute of Physics, HR-10000 Zagreb, Croatia*

*email: mar331@pitt.edu, petek@pitt.edu*

# S1 2PP spectroscopy of Cu(111)

The projected surface band structures of Ag(111) and Cu(111) are qualitatively similar: The SS state as well as the $n = 1$ IP state are within their band gaps while the higher lying IP states ($n \geq 2$) are resonant with the $U_{sp}$-band [1-7]. Besides minor differences in binding energies, $k_\parallel$-dispersions and electron dynamics of the electronic states, both surfaces mainly differ in their dielectric functions and thus, most pertinently, their bulk plasmon frequencies [8]. While for Ag, the bulk plasmon frequency is smaller than the work function ($\Phi \approx 4.5$ eV vs. $\hbar\omega_p \approx 3.9$ eV), the bulk plasmon frequency of Cu is not well-defined, but has been estimated to be in ~20 eV range [8,9].

In Fig. S1 and Fig. S2, we show 2PP spectroscopy of the clean Cu(111) surface as excited in a photon energy range of 2.88 eV < $\hbar\omega$ < 4.62 eV where we can contrast the characteristics of both the SP transition as well as the IP ← SS resonance, with those of Ag(111).

**sp-band intensity**

In Fig. S1(a), we show an $E_f(k_\parallel)$ distribution of photoelectrons recorded in 2PP from Cu(111) upon excitation with $\hbar\omega = 3.41$ eV photons. The SP transition and the SS state are detected in resonant and non-resonant coherent 2-photon excitation, respectively [the excitation diagram as a function of $k_\perp$ is presented in Fig. S1(c)]. In Fig. S1(b), we show 2PP spectra for $k_\parallel = 0$ for $\hbar\omega = 2.88$ to 4.22 eV, which are in agreement with literature [2]. We first focus on the 2-photon SP transition, which has the following properties: (i) it is detected in a narrow photon energy range, 3.3 eV < $\hbar\omega$ < 3.5 eV; (ii) it has a parabolic $k_\parallel$-dispersion; (iii) it is superimposed on d-band photoemission for $\hbar\omega$ > 3.5 eV; and (iv) its photoemission intensity is substantially weaker than that of the SS state.

Although Cu(111) and Ag(111) have similar single particle band structures, their SP transitions appear with dramatically different relative intensities. Whereas for Cu(111) the SP transition is hardly detected in 2PP, for Ag(111) it dominates 2PP spectra for $\hbar\omega$ < 3.9 eV [Fig. 2(a), Fig 3(b), Fig. 4(b)]. This difference can be attributed to their different dielectric functions. As outlined in the main text, for Ag(111), the resonant 2-photon SP transition gains intensity by enhancement of the local field from the multipole surface plasmon response. By contrast, on Cu(111), ENZ and thus strong plasmonic field enhancements cannot be expected in the energy range where the SP transition is observed (3.3 eV < $\hbar\omega$ < 3.5 eV), as these are expected at a much higher energy [8]. The SP transition for Ag(111) has been observed in 1PP for $\hbar\omega$ ~ 6-10 eV, where a plasmonic field enhancement is not expected [10]; by contrast to 2PP spectra, in 1PP spectra, the SP transition is not strongly $\hbar\omega$ dependent, and it strength is comparable or weaker than that of SS. We thus conclude, that the strong $\hbar\omega$ dependence of the 2-photon SP transition in Ag(111) can be attributed to the local field enhancement due to the plasmonic screening of the optical field below $\hbar\omega_p$.

**IP ← SS resonance**

In Fig. S2, we show 2PP spectra of Cu(111) when tuning the photon energy through the SS ← $n$ = 1 IP resonance, the photoemission intensity is enhanced for selected $k_\parallel$ due to the different $k_\parallel$-dispersions of the coupled states. Most relevant for this work is the following observation: When comparing the photon energy dependent 2PP data of Cu(111) in Fig. S2 with the 2PP data of Ag(111) in Fig. 2 and Fig. 3 of the main text, the surface states behave similarly, *i.e.,* their photon energy dependent spectroscopy is mainly dominated by the $n$ = 1 IP ← SS resonance. Because for Cu(111) surface, plasmonic effects are not expected [8,9], the similar behavior of the same transition on Ag(111) suggest that plasmonic effects do not strongly affect its surface state 2PP processes.

## S2 2PP and 3PP spectroscopy of Ag(111)

In 1985, Giesen *et al.* [11] reported 2PP spectroscopy of Ag(111) using tunable nanosecond dye laser pulse excitation and reported similar spectroscopic features as discussed in this Letter. As we outline in the following, we do not concur with some of their assignments. Giesen *et al.* did not consider the role of plasmonic excitations, but mainly the resonance condition between the SS state and the $n = 1$ IP states when exciting with $\hbar\omega \geq \hbar\omega_p \approx 3.9$ eV photons.

Specifically, Giesen *et al.* assigned the $2\hbar\omega_p$-feature to energy-pooling, an Auger-like process where a large density of electrons excited to the $n=1$ IP state can decay by impact ionization where two IP state electrons scatter causing one to decay to $E_F$ and the other to acquire the initial energy of both. This would cause the ionized electron to appear at twice the $n = 1$ IP state energy (at $k_\parallel=0$ Å$^{-1}$), which is nearly degenerate with $E_f$ of the $2\hbar\omega_p$-feature. We can exclude the Auger-like decay process based on the following arguments and experimental results. (i) Energy pooling, as a two-electron process, should be strongly dependent on the $n = 1$ IP state population. In our work, we excite the IP state with femtosecond laser pulses, whereas Giesen *et al.* worked with nanosecond lasers. The IP state population should depend on both the laser pulse duration and the IP state lifetime. Because $n = 1$ IP state lifetime is $\approx 30$ fs [4], which was not known to Giesen *et al.*, the energy pooling process should be far more efficient with our fs excitation source, but this is not the case. Because the $2\hbar\omega_p$-feature is detected with qualitatively comparable relative intensity to the $n = 1$ IP state, we can rule out energy pooling, as being responsible for the $2\hbar\omega_p$-feature. (ii) In Fig. S3, we show the $n = 1$ IP state region in 2PP spectra of Ag(111) excited with $\hbar\omega = 4.22$ eV together with a three-photon photoemission (3PP) excited with $\hbar\omega = 2.12$ eV. As discussed in the main text, for $\hbar\omega \geq \hbar\omega_p$, the $2\hbar\omega_p$-feature is clearly resolved and appears in the 2PP spectrum. Energy-pooling should not depend on the excitation pathway, i.e. whether the IP state is excited by 1- or 2-photons and thus it should be observed when exciting with $\hbar\omega$ = 2.12 eV, because we can generate a large density in the $n$ = 1 IP state. As can be clearly seen in Fig. S3(b) and Fig. S3(c), however, although the $n$ = 1 IP state is strongly populated by two photon excitation with $\hbar\omega$ = 2.12 eV, the $2\hbar\omega_p$-feature is not observed when $\hbar\omega < \hbar\omega_p$; therefore, the assignment of energy pooling to the $2\hbar\omega_p$-feature can be excluded. (iii) For energy pooling to be $\hbar\omega$ independent, energy relaxation must occur from $k_{||} \neq 0$ Å$^{-1}$ states to $k_{||} = 0$ Å$^{-1}$ before the Auger process occurs. Such fast intraband relaxation is not detected in IP states. (iv) If energy pooling occurs in Ag(111), it could also occur in Cu(111); there is no evidence for that in, for example, Fig. S2. The argument against energy pooling is further supported by observation of above-threshold photoemission (ATP) of the IP state with $\hbar\omega$ = 2.12 eV: The ATP signal is typically a factor 10 to 10$^3$ weaker than the lower-order photoemission process [12-14]. Because our experiment is sensitive to ATP electrons, as from the $n$ = 1 IP state, it should also be sensitive to electrons from energy-pooling. Further arguments are given in section S3.

## S3 Molecular–coverage dependent 2PP spectra of Ag(111)

The electronic structure and the electron dynamics at molecule-metal interfaces are intensively studied with 2PP [15,16]. Here, we make use of 3,4,9,10-perylene-tetracarboxylic acid dianhydride (PTCDA) adsorption on Ag(111) in order to eliminate the possibility of $n$ = 1 IP ← SS state resonance from contributing to the $2\hbar\omega_p$-feature, and thus to exclude the Auger-like decay process proposed by Giesen *et al.* [11].

Upon adsorption of a few-layer PTCDA on Ag(111), the SS of the clean Ag(111) surface becomes an unoccupied Shockley-type resonance of the metal-organic interface [16,17]. In Fig. S4, we show PTCDA coverage dependent 2PP data ($\hbar\omega$ = 4.58 eV, $k_\parallel$ = 0, 300 K). With increasing coverage, the Shockley-type interface state (IS) is formed and detected in 2PP, while the intensity of SS state of the clean surface drops significantly. For highest coverage ($\geq$ 1ML), the SS state signal has disappeared, because it has become the IS state [17]; however, the $2\hbar\omega_p$-feature is still visible in the data. The $n$ = 1 IP state peak is still observable, but that is because the IP state can exist on the organic/vacuum interface. We make use of the coverage dependent 2PP data to conclude that the SS state and the $2\hbar\omega_p$-feature are independent photoemission spectral features. This observation strongly contradicts the attribution of the $2\hbar\omega_p$-feature to an Auger-like decay process proposed in Ref. [11].

## S4 Rb/Ag(111) - coverage dependent 1PP spectra excited with a Hg-lamp

We report the detection of electrons that appear to be excited by decay of bulk plasmons in Ag(111). For the pristine Ag(111) surface, single particle electrons excited by one bulk plasmon ($\hbar\omega_p \approx$ 3.8-3.9 eV) do not have sufficient energy to overcome the work function to be photoemitted. Instead, as discussed in main text, in 2PP electrons excited by decay of two bulk plasmons are detected at a final state energy of $2\hbar\omega_p \approx$ 7.75 eV. If this excitation pathway is feasible, we also might expect electrons to be excited to $E_f$ = $1\hbar\omega_p \approx$ 3.8 eV, and be emitted if the work function were sufficiently low.

To test this hypotheses, we decrease the work function of Ag(111) by deposition of sub-monolayer coverage of Rubidium. 1PP and 2PP spectroscopy of alkali atom adsorption on noble metal surfaces is well documented in literature [18-22], notably as well as for alkali atoms on Ag(111) [23]. In Fig. S5, we show Rb-atom coverage dependent 1PP spectra excited with $\hbar\omega$ = 4.86 eV photons from an Hg-lamp. The Rb coverage of <0.1 monolayer is increased to achieve work function reduction comparable to similar work of Barman *et. al.* (Fig. 5 in Ref. [22], $\hbar\omega \approx 5$ eV). For the clean Ag(111) surface, only the SS state is detected in 1PP [Fig. S5 (a) (left) and Fig. S5 (b)]. Gradually increasing Rb coverage decreases the work function and allows the resonant 1-photon SP transition to be detected [Fig. S5 (a) (right), excitation diagram in Fig. S5 (c)]. Most importantly, we observe a photoemission spectral element ("peak") at $E_f \approx 3.7$ eV, $1\hbar\omega_p$, which shows a broad line shape both in energy as well as in momentum space. Based on the known band structure of Ag(111) [1,3,6,24], this cannot be excited by single-particle excitations, because there is no initial or final single-particle state that could be excited in 1PP at this energy. Instead, we attribute the $1\hbar\omega_p \approx 3.7$ eV feature to electrons being excited by the decay of one bulk plasmon quantum. This photoemission signal must involve a decay of a collective state, because there are no single-particle initial or final states within the bulk band gap of Ag(111) at $k_\parallel = 0$ that could be the origin of this emission (for details see main text). Moreover, the broad line shape is not compatible with single particle features in 1PP or 2PP spectra of Ag(111). We note that the 1PP spectra of Barman *et al.* also detected a weak shoulder at ~3.7 eV probably of the same origin [22]. We thus conclude that the decay of bulk plasmon quanta can preferentially excite electrons from $E_F$.

We note, that the Hg-lamp emission incudes a line at $\hbar\omega \approx 3.7$ eV (factor 15 weaker when compared to $\hbar\omega \approx 4.86$ eV), which could lead to 1PP of SS state electrons to $E_f \approx 3.7$ eV. We exclude this as a possible contribution to the peak at $E_f \approx 3.7$ eV, however, because its line shape and $k_\parallel$-dispersion significantly differ from those of the SS state [cf. Fig. S5 (a)].

## S5 Temperature dependent 2PP spectroscopy of Ag(111)

To further determine origin of the $2\hbar\omega_p \approx 7.75$ eV feature, we investigate its temperature dependence in 2PP spectra. The motivation for such measurement is that if the emission involves states close to $E_F$, then the 2PP spectrum should be sensitive to temperature because it will be influenced by the Fermi-Dirac distribution, and electron-phonon interaction. By contrast, if it involves states far from $E_F$, the spectra should be independent of temperature, because electron occupations will be 0 or 1, and the lifetimes will be determined by electron-electron interaction. One caveat is that long lived surface states far from $E_F$ could also be significantly influenced by electron-phonon interaction [25].

Figure S6 shows temperature dependent 2PP data of Ag(111) excited with 4.22 eV photons ($k_\parallel = 0$ Å$^{-1}$). For 90 and 300 K sample temperature, the $n = 1$ IP and the SS states as well as the $2\hbar\omega_p$-feature are resolved, however, as the spectra clearly indicate, both the $2\hbar\omega_p$-feature as well as the $n = 1$ IP state are temperature dependent: At 300 K, their linewidths are significantly broadened as would be expected for electrons, which are excited from the Fermi level or sensitive to electron-phonon interaction. The broader width and larger temperature sensitivity of the $2\hbar\omega_p$-feature may be related to the canonical $4k_BT$ (100 meV at 300 K) width of Fermi-Dirac distribution.

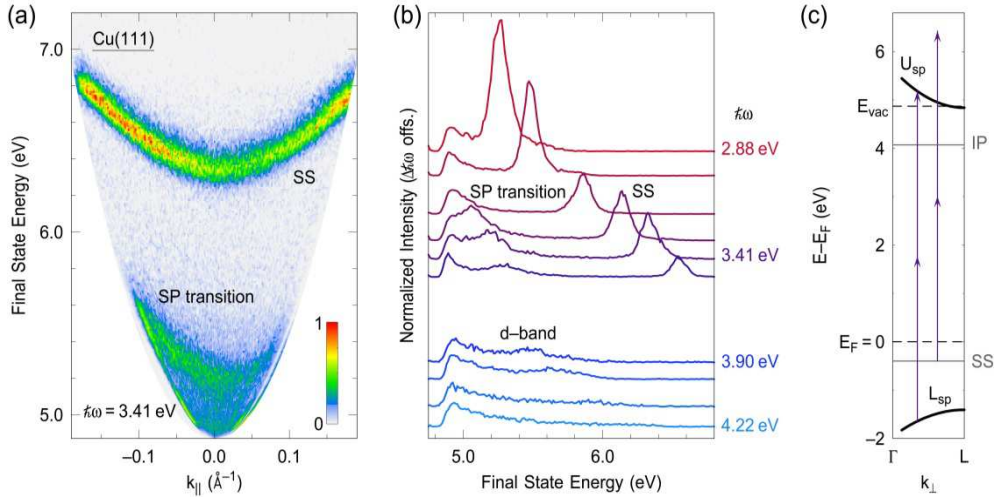

**Figure S1** | (a) $E_f(k_\parallel)$-distribution of Cu(111) as excited with 3.41 eV photons, the spectrum is composed of the coherent non-resonant 2-photon SS state excitation and the resonant SP transition [the excitation diagram as a function of $k_\perp$ is given in (c)]. (b) Line profiles taken at $k_\parallel = 0$ in a photon energy range between 2.88 and 4.22 eV. The spectra are normalized at the work function edge and shifted on the vertical axis by $\Delta\hbar\omega$.

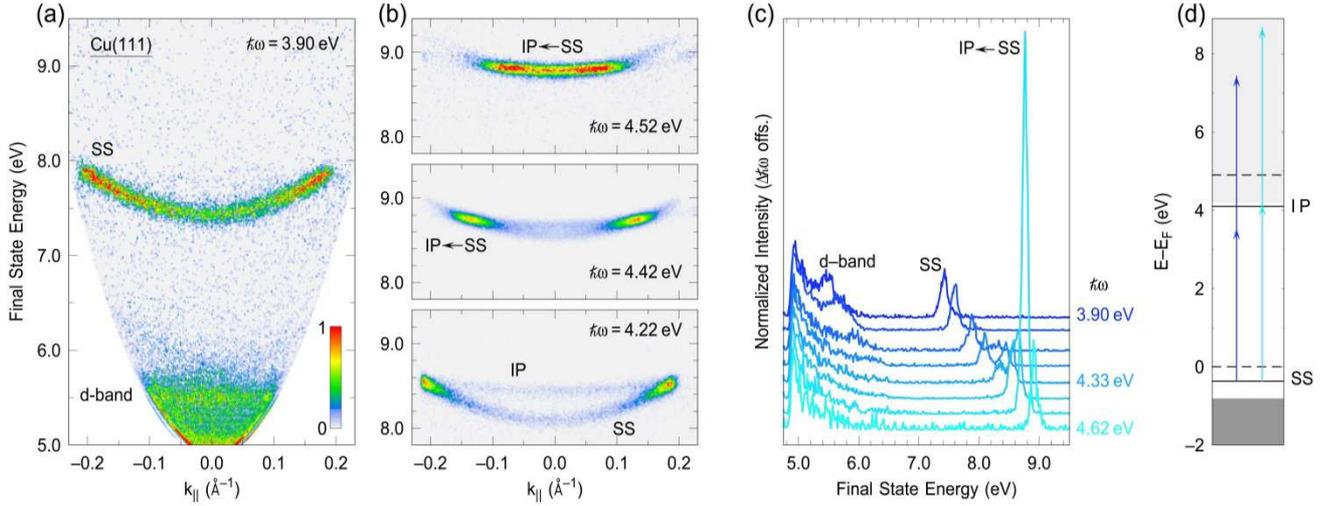

**Figure S2** | Energy- and $k_\parallel$-resolved 2PP spectra of the Cu(111) surface for photon energies between 3.90 and 4.62 eV, each color table is scaled independently. (a) For $\hbar\omega = 3.90$ eV, the SS state as well as the d-band are detected. (b) When increasing the photon energy from 4.2 to 4.5 eV, an intensity enhancement due to the $n = 1$ IP ← SS resonance is observed for different $k_\parallel$, because of their different dispersions. (c) Line profiles of Cu(111) spectra taken at $k_\parallel = 0$; the spectra are normalized at the work function edge and shifted on the vertical axis by $\Delta\hbar\omega$. The 2PP spectra of Cu probe the very similar single-particle band structure, but do not exhibit plasmonic responses found for Ag(111). (d) Excitation diagram for $k_\parallel = 0$, 2PP of the SS state at $\hbar\omega = 3.90$ eV (blue) as well 2PP in the IP ← SS resonance (cyan) are indicated by the arrows.

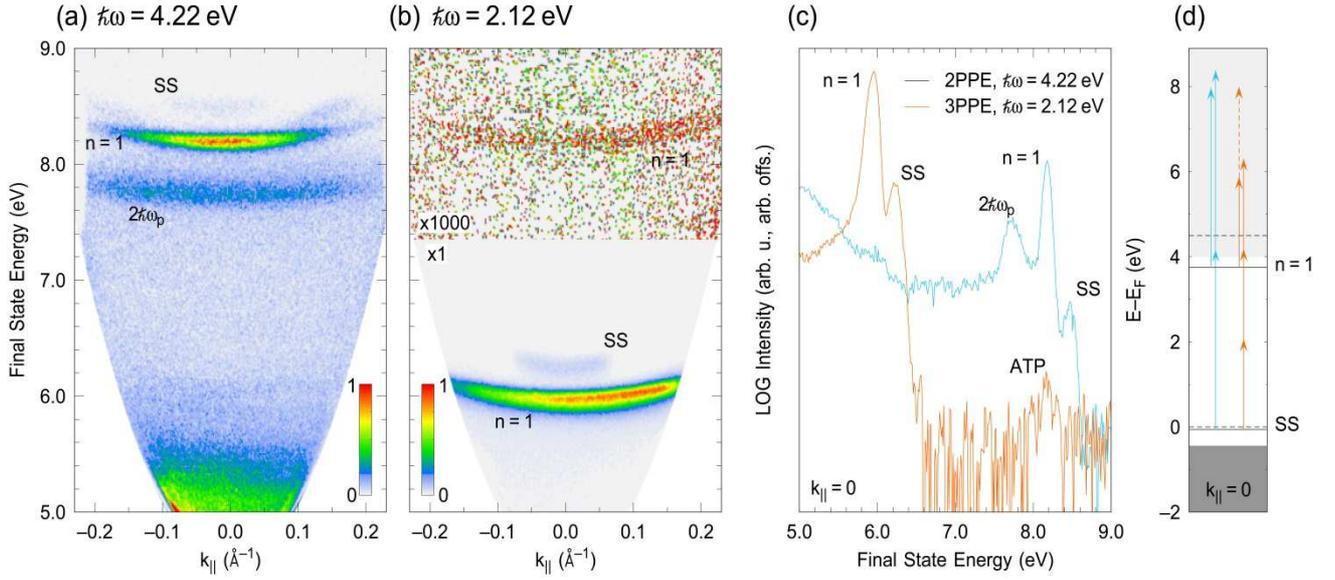

**Figure S3** | Energy- and $k_\parallel$-resolved (a) 2PP and (b) 3PP spectra when exciting with $\hbar\omega = 4.22$ eV and $\hbar\omega = 2.12$ eV, respectively (the color tables are scaled independently). In (b), the color scale is enhanced by a factor 1000 for $E_f > 7.3$ eV to show the ATP signal. (c) Line profiles taken at $k_\parallel = 0$ are shown on a logarithmic scale of intensity. (d) Excitation diagram at $k_\parallel = 0$, the cyan and the orange arrows indicate the 2-photon and 3-photon excitation processes from SS, respectively; the dashed arrow indicates ATP. For $\hbar\omega = 4.22$ eV, the $n = 1$ IP state, the SS state as well as the $2\hbar\omega_p$-feature are resolved in 2PP. For $\hbar\omega = 2.12$ eV, the $n = 1$ IP state, the SS state are resolved in 3PP; a replica of the $n = 1$ IP state is observed in ATP.

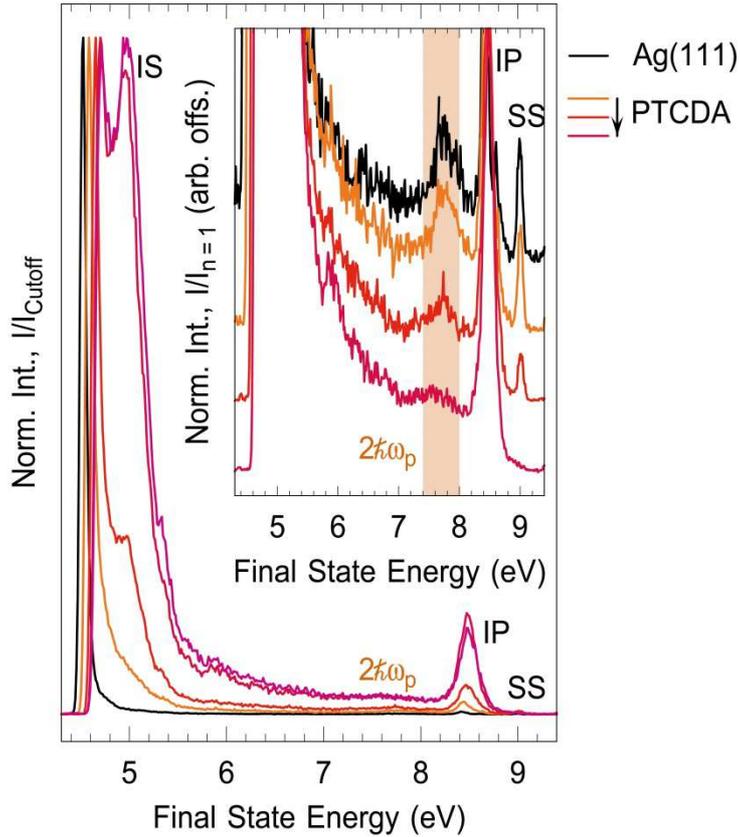

**Figure S4 |** 2PP spectra for different PTCDA coverages ($\hbar\omega$ = 4.58 eV, $k_{\parallel}$ = 0, 300 K). The spectra are normalized at the work function edge. In the inset, the signal is normalized on the $n$ = 1 IP state intensity and offset by an arbitrary amount on the vertical axis. With increasing PTCDA coverage, the initially occupied SS state disappears and the unoccupied Shockley-type metal-organic interface state (IS) is formed. The SS state intensity drops significantly faster than for the $2\hbar\omega_p$-feature.

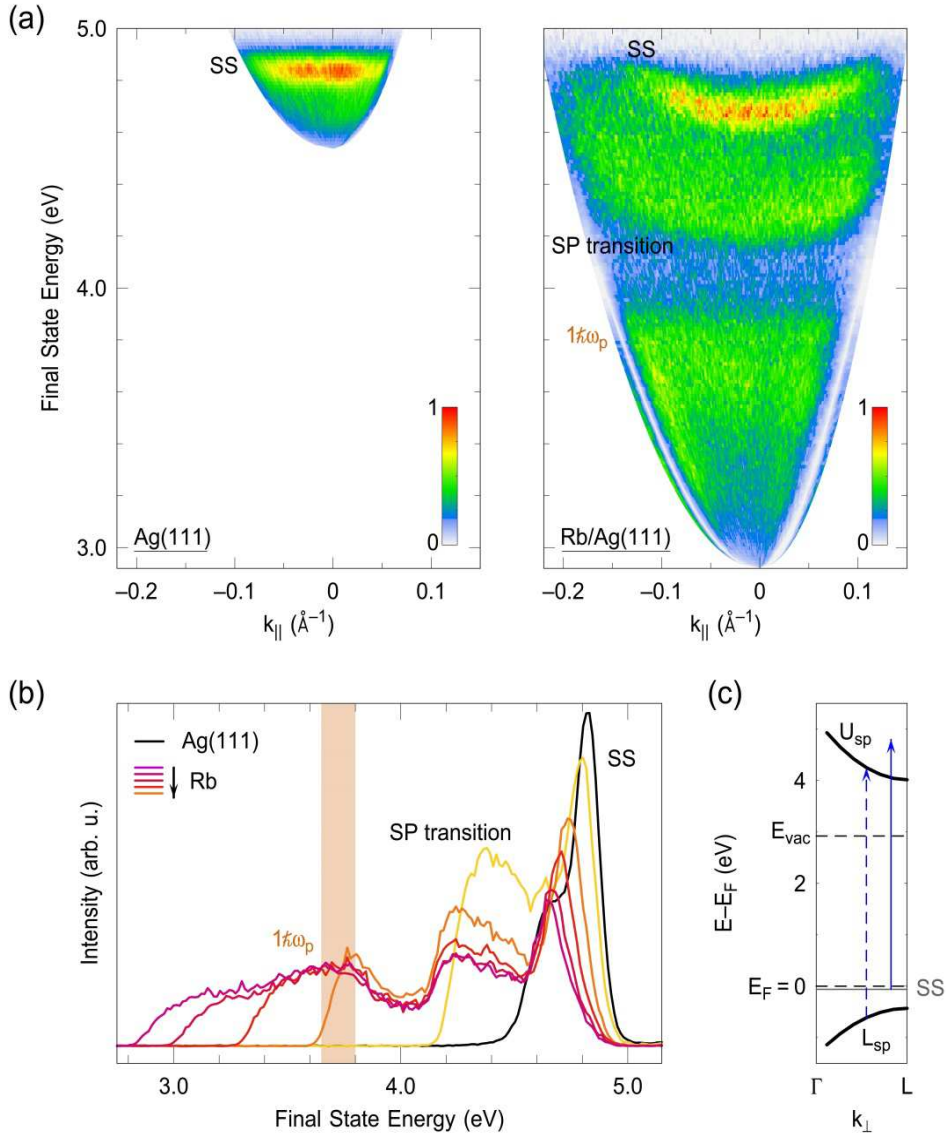

**Figure S5** | (a) $E_f(k_\parallel)$-distributions as excited with a Hg-lamp ($\hbar\omega = 4.86$ eV) of the pristine Ag(111) (left) and the Rb/Ag(111) (right) surface. The photoemission spectral elements are labelled in the figure. (b) Energy-resolved 1PP spectra of Rb/Ag(111) for $k_\parallel = 0$; the Rb coverage is increased continuously to lower the work function. For the pristine Ag(111) surface (black spectrum), the SS state is resolved. As the work function decreases from ≈4.5 eV (clean surface, black spectrum) to ≈3.1 eV (purple spectrum), the resonant 1-photon transition between the lower, $L_{sp}$, and the upper, $U_{sp}$, sp-band becomes apparent at a final state energy of $E_f \approx 4.2$ eV. At a final state energy of $E_f \approx 3.7$ eV = $1\hbar\omega_p$, electrons as emitted by the decay of one bulk plasmon are detected. The $1\hbar\omega_p$–features has a broad line shape in energy and momentum space. (c) $k_\perp$-resolved band structure in Γ-L direction, the 1-photon SP transition (dashed line) as well as the 1-photon SS state photoexcitation (solid line) are sketched ($E_{vac} \approx 3.1$ eV).

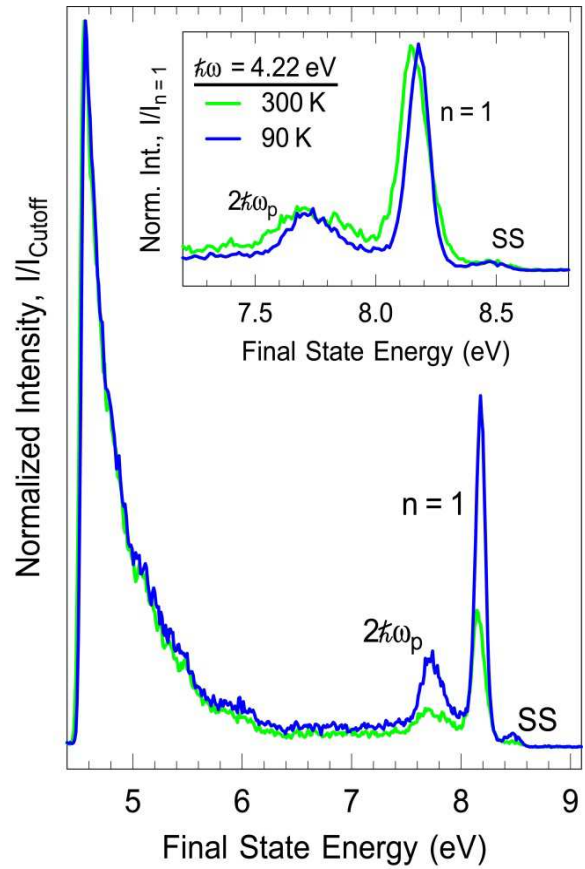

**Figure S6** | Temperature dependent 2PP spectra taken with 4.22 eV photons ($k_{\parallel} = 0$). The $n = 1$ IP state and the SS state as well as the $2\hbar\omega_p$-feature are resolved at 90 (blue) and 300 K (green) sample temperatures; these features become sharper and more intense at 90 K. The spectra are normalized at the work function edge. In the inset, the signal is normalized on the $n = 1$ IP state intensity.

# References.